\documentclass[aps,pra,preprint,showpacs,amsmath,amssymb,floatfix,unsortedaddress,superscriptaddress,10pt]{revtex4-1} 

\usepackage[utf8]{inputenc}  
\usepackage{graphicx}        
\usepackage{bm}              
\usepackage{subfigure}
\usepackage{xcolor}
\usepackage[colorlinks=true,citecolor=blue,urlcolor=blue]{hyperref}

\begin{document}


\title{Single attosecond pulse from terahertz assisted high-order harmonic generation}

\author{Emeric Balogh}
\affiliation{Department of Optics and Quantum Electronics, University of Szeged, H-6701 Szeged, Hungary}

\author{Katalin Kovacs}
\affiliation{Department of Optics and Quantum Electronics, University of Szeged, H-6701 Szeged, Hungary}
\affiliation{National Institute for R\&D of Isotopic and Molecular Technologies, RO-400293 Cluj-Napoca, Romania}

\author{Peter Dombi}
\affiliation{Research Institute for Solid State Physics and Optics, H-1525 Budapest, Hungary}

\author{Jozsef A. Fulop}
\affiliation{Department of Experimental Physics, University of P\'{e}cs, H-7624 P\'{e}cs, Hungary}

\author{Gyozo Farkas}
\affiliation{Research Institute for Solid State Physics and Optics, H-1525 Budapest, Hungary}

\author{Janos Hebling}
\affiliation{Department of Experimental Physics, University of P\'{e}cs, H-7624 P\'{e}cs, Hungary}

\author{Valer Tosa}
\affiliation{National Institute for R\&D of Isotopic and Molecular Technologies, RO-400293 Cluj-Napoca, Romania}

\author{Katalin Varju}
\affiliation{HAS Research Group on Laser Physics, University of Szeged, H-6701 Szeged, Hungary}

\date{\today}

\begin{abstract}
High-order harmonic generation by few-cycle 800 nm laser pulses in neon gas in the presence of a strong terahertz (THz) field is investigated numerically with propagation effects taken into account.
Our calculations show that the combination of THz fields with up to 12 fs laser pulses can be an effective gating technique to generate single attosecond pulses.
We show that in the presence of the strong THz field only a single attosecond burst can be phase matched, whereas radiation emitted during other half-cycles disappears during propagation.
The cutoff is extended and a wide supercontinuum appears in the near-field spectra, extending the available spectral width for isolated attosecond pulse generation from 23 to 93 eV.
We demonstrate that phase matching effects are responsible for the generation of isolated attosecond pulses, even in conditions when single atom response yields an attosecond pulse train.
\end{abstract}

\maketitle


\section{Introduction}
The shortest - attosecond - light pulses available today are produced by high-order harmonic generation (HHG) \cite{farkas, PRL.77.1234, krauszrevmod}, a process in which an electron extracted by an intense laser field from an atom is accelerated as a free particle and, upon recombination with the parent atom, releases its kinetic energy in the form of a single photon. 
In this way near-infrared (NIR) laser pulses of 800 nm wavelength can provide a broad spectral plateau of extreme ultraviolet (XUV) radiation ending in a cutoff.
The minimum pulse duration is determined by the achievable bandwidth (i.e. the position of the cutoff) and the chirp of the produced radiation.
The cutoff scales with the intensity, but the extension of the cutoff by increase in the laser intensity is limited by the depletion and phase matching problems arising in the ionized medium.
An alternative method demonstrated to produce higher harmonic orders is by using longer pump pulse wavelength, with the disadvantage of decreased efficiency \cite{tate}.

For a monochromatic field an attosecond pulse is created every half cycle of the generating laser pulse \cite{Paul01062001}.
To produce single attosecond pulses (SAPs) for clean pump-probe studies, very short - few cycle - laser pulses have to be used \cite{Drescher09032001, nature_AttoMetro, krauszrevmod}.
Such short laser pulses are hard to obtain reliably, therefore alternative ways have been developed to gate high-order harmonic generation, restricting the XUV production to a single half-cycle and thus leading to SAPs.
The first technique to provide SAP generation used intensity gating, selecting the cutoff region of the radiation produced by a few-cycle pulse, with only a half-cycle contribution \cite{Drescher09032001}.
Polarization gating proved to be an efficient technique for achieving SAP production with long laser pulses \cite{timegate,sola_polgate,sola_polgate2}.
Ionization gating is another way to produce SAP, based on the principle that by use of very high laser intensities the medium is depleted on the rising edge of the pulse \cite{sekikawa_iongate,ferrarinphys}.

Another way is two color gating, which is obtained by combining two or more laser pulses having different wavelengths and usually different intensities \cite{siedschlag_twocolorSAP}.
The electromagnetic field of the stronger laser pulse (commonly from a Ti:Sapphire laser at 800 nm wavelength) is called the driving field, while that of the weaker one is the control field.
The role of the latter usually is to control the movement of the electrons while away from the nucleus, preventing or helping recombination, or to control the exact time of the ionization by which the duration, intensity and also the chirp of the produced harmonic pulse can be affected.
The combination of a short laser pulse with an even shorter ultraviolet or extreme ultraviolet pulse was demonstrated to be an efficient technique to control the time of ionization \cite{GSirxuv,johnssonprl}.
The wavelength of the control field can also be longer than that of the driving field as demonstrated by Vozzi et al. \cite{milano800_1600} and Calegari et al. \cite{Calegari09}.

The fast development of terahertz (THz) pulse production by difference frequency generation made it possible to obtain extremely strong, long wavelength fields, the electric field reaching 100 MV/cm \cite{sell}.
The method described by Sell et al. \cite{sell} also enables the tuning of the resulting field's wavelength in a wide range, down to the mid-infrared domain.
The strength of the electric field obtained by this method is already comparable with that of the laser pulses traditionally used when high-order harmonics are produced in gases, therefore it may alter the HHG process considerably.

The effects of using THz and static electric fields in HHG at single atom level have been investigated in several papers \cite{minqi, milosevic, hong, yang, xiaPhysRevA.81.043420, varjuthz}. 
In this work, we further investigate the idea at the macroscopic level, by considering all the important effects arising during the propagation of electromagnetic fields, simulating a scenario with realistic focusing geometry.
The strong field approximation (SFA) \cite{lewenstein} is used to calculate the dipole radiation from individual atoms, the non-adiabatic saddle point approximation \cite{sansone} to analyze electron trajectories, and propagation effects in the laser, THz, and harmonic fields are taken into account when the wave equation is solved for them \cite{priori,tosamodel}.
We calculate harmonic generation in Ne atoms, with few cycle, near infrared pulses having a peak intensity of 6-10$*10^{14}$ W/cm$^2$ and 800 nm wavelength.
The peak amplitude of the THz field is 100 MV/cm, which is the highest field strength currently available \cite{sell}.

This paper is organized as follows: In \autoref{sec:methods} the model used to simulate high-order harmonic generation in a macroscopic medium is presented.
In \autoref{sec:setup} the main parameters of the model are presented, followed by the results in the single atom case, and also the generated harmonic bursts arriving at the exit of the interaction region.
The last part of \autoref{sec:setup} focuses on the interpretation of the results by analyzing phase matching effects arising during the process.
In \autoref{sec:discussion} the prospects and limits of the method are discussed, followed by the conclusions in \autoref{sec:conclusions}.

\section{Model}
\label{sec:methods}

As mentioned before, the process of high-order harmonic generation is well understood using the three step model \cite{corkum}, which pictures HHG as the result of (1) tunneling ionization, (2) free electron movement in the laser field, and (3) recombination.
Since the electric field of the laser pulse is comparable with the Coulomb field of the atom, at certain phases of the laser field the outermost electron may tunnel into the continuum.
In the second step the free electron is driven by the sinusoidal electric force of the laser, and first accelerated away from the nucleus, then pulled back toward the parent ion on the change of the sign of the electric field.
Revisiting the parent ion the electron may recombine and release its energy in the form of a high energy photon.

The exact treatment for describing the emitted radiation from the interaction of the laser pulse with a single atom is to solve the time dependent Schr\"{o}dinger equation.
Use of the SFA of Lewenstein et al. \cite{lewenstein} enables easier numerical treatment and interpretation of the results while still reproducing the main aspects of the process.
The analytical simplification to the Schr\"{o}dinger equation, proven to be valid in conditions under which high-order harmonics are generated, gives a simpler formula for the time dependent dipole moment.
The method is widely used when the process of HHG is analyzed, especially if macroscopic effects are taken into account which needs the calculation of the single atom response to be done at least thousands of times making the use of the Schr\"{o}dinger equation impractical.

The Lewenstein integral for the time dependent dipole moment reads:
\begin{widetext}
\begin{align}
d(t) &= 2Re \Big\{ i\int_{-\infty}^t dt' \Big( \frac{\pi}{\epsilon+i(t-t')/2} \Big)^{3/2} d^{\ast}[p_{st}(t',t)-A(t)]d[p_{st}(t',t)-A(t')]exp[-iS_{st}(t',t)]E(t')\Big\}
\label{eq:lew}
\end{align}
\end{widetext}
where $E(t)$ and $A(t)$ denote the time dependent electric field and vector potential of the laser pulse, d(p) denotes the atomic dipole matrix element for the bound-free transition, $\epsilon$ being a small positive number to remove the divergence at $t=t'$, while $p_{st}$ is the stationary point of the canonical momentum, and $S(t',t)$ is the quasiclassical action defined as:
\begin{align}
S(t',t) &= i\int_{t'}^t \Bigg( \frac{(p(t'')-A(t''))^{2}}{2}+I_p \Bigg) dt'',
\label{eq:sp}
\end{align}
$I_p$ being the ionization potential.

The harmonic emission spectrum is obtained as the Fourier transform of the time dependent dipole moment, which is filtered from low order harmonics for which the formula is not accurate.
The temporal shape of the produced attosecond pulses is calculated by inverse Fourier transform of the harmonic field in a given spectral range.

In \autoref{eq:lew} the dipole moment is calculated taking into account all electron trajectories, on which the electron leaves the atom before time $t$ and recombines with the parent ion exactly at $t$.
However, it was found that only a few of these infinite number of trajectories are actually relevant (the stationary points of the phase), therefore by selecting only these, and approximating the value of the integral using the saddle point approximation \cite{lewenstein, sansone, sansone09, katus} further information can be obtained that makes possible the trajectory analysis.
The first two trajectories with the shortest travel time are distinguished as the most important ones, and are called the short and long trajectories.
Those having a travel time more than one optical cycle are usually called superlong trajectories.

We found that the effect of superlong trajectories can be important, at least at single atom level, which is not surprising since Tate et al. described it \cite{tate}.
Therefore the Lewenstein integral was calculated for a time period as long as $3$ optical cycles of the fundamental NIR field.
Further increase in the length of the time domain causes no considerable change in the resulting spectra, or in the synthesized harmonic pulses.

In order to include the effect of propagation on the laser, THz and harmonic fields the corresponding wave equations in paraxial approximation were solved for all of them.
The detailed description of the fundamentals of this method has been given by Priori et al. \cite{priori} taking into account the ionization and plasma dispersion. 
This has been completed by including the effects of absorption, dispersion on atoms and the optical Kerr effect by Takahashi et al. \cite{tosamodel}.
In our case the propagation of the laser and THz fields cannot be completely separated, since the optical Kerr effect and the ionization depend on the total electric field to which the atoms are exposed, therefore related quantities are calculated for the combined field before solving the wave equation.
In this study the ADK model is used for calculating the ionization rate \cite{adk}.
The trajectory analysis for arbitrary laser fields using saddle point approximation, which is also included, was described in \cite{katus}.

\section{Configuration and Results}
\label{sec:setup}

In the configuration we assumed, the laser pulse has a central wavelength of 800 nm while the wavelength of THz field is 8 $\mu$m (corresponding to 37.5 THz).
In the two main cases the peak intensity of the IR field is $6*10^{14}$ W/cm$^2$ produced by 0.3 mJ, 5.2 fs (or 0.47 mJ, 8 fs) pulses of a beam of 2 mm diameter, focused by a mirror of 0.6 m focal distance, resulting in a beam waist of 76 $\mu$m in the focus with a Rayleigh range of 22.9 mm.
The peak amplitude of the THz field is 100 MV/cm, a value obtained experimentally by Sell et al. \cite{sell}.

Both the IR and THz pulses are treated as Gaussian beams, focused at the same spot. The 1 mm long gas cell containing neon gas with a pressure of 15 Torr is placed right after the focus.
To have the best spatial overlap between the two pulses, the THz field is focused to have the same 76 $\mu$m beam waist, resulting in a Rayleigh range of 2.29 mm.
Both pulses propagate in the same direction, having parallel linear polarization, and are synchronized so their peaks overlap at the focus.
When the limits of the method or certain aspects of the process were tested, some of these parameters were changed, as specified later.

\subsection{Terahertz and laser field propagation}
\label{sec:fieldprop}

High-order harmonic generation relies on the process of recapturing the optically ionized electron after its propagation in the laser field. 
The characteristics of the radiation are ultimately dependent on the electric field to which the atoms are exposed.
During the propagation through the gas cell, several linear and nonlinear effects may disturb the laser and THz pulses, such as absorption, dispersion on atoms, the optical Kerr effect, and dispersion on electrons (plasma dispersion).
The effect of the plasma dispersion is particularly important in the case of low frequency fields like the THz field, because this effect scales with $(\omega_p/\omega)^2$ (where $\omega_p$ is the angular plasma frequency, and $\omega$ is the angular frequency of the propagated field).

The results show that under the conditions we proposed, the IR field is almost unchanged during the propagation, because the ionization is quite low, and the pulse's Rayleigh length is much longer than the gas cell.
At the exit of the interaction region (after 1 mm propagation) the peak of the pulse on axis is ahead by ~16 as, which is mostly caused by the Gouy phase shift that yields an 18.5 as change in the same direction.

\begin{figure}[htb!]%
	\subfigure{
	\includegraphics{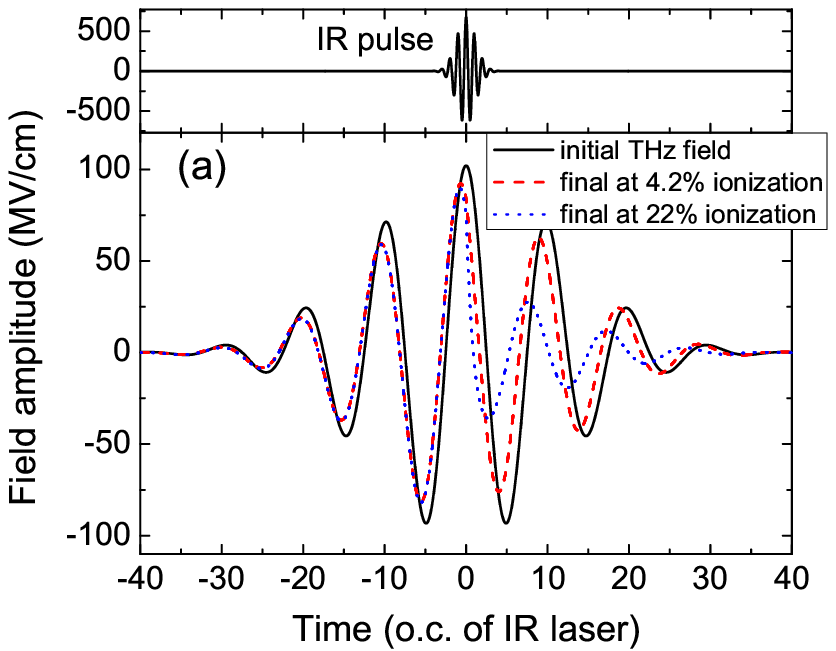}%
	\label{Fig:01a}}
	\subfigure{
	\includegraphics{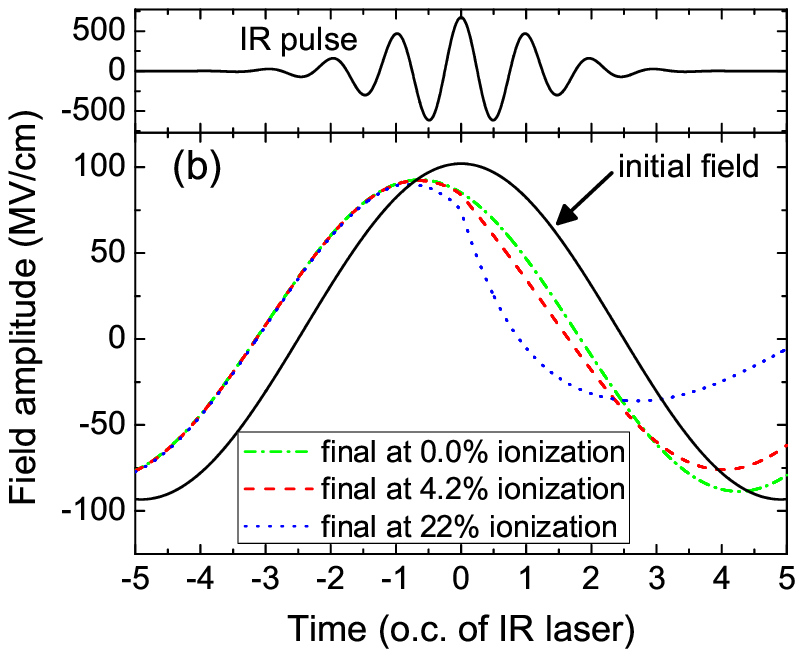}%
	\label{Fig:01b}}
\caption{(Color online) Electric field of the THz pulse at the focus (black solid lines) and after 1 mm propagation in the presence of a 5.2 fs laser pulse, with a peak intensity of $6*10^{14}$ W/cm$^2$ (red dashed line) and 10$^{15}$ W/cm$^2$ (blue dotted line). The upper insets show the incident laser field for the $6*10^{14}$ W/cm$^2$ case. (b) Central portion of the THz pulse's electric field in the focus (black solid line), and after 1 mm propagation in a medium with 0\% (green dash-dotted line), 4.2\% (red dashed line), and 22\% (blue dotted line) ionization, indicating the strong distortion of the THz field due to ionization right after the peak of the IR pulse.
}%
\label{Fig:01}%
\end{figure}

As can be seen in \autoref{Fig:01a} the macroscopic effects are much more evident in the case of the THz field.
Because of the 2.29 mm Rayleigh range, the Gouy phase shift is 0.4 radians at 1 mm from the focus, which yields a 1.74 fs shift of the pulse's peak, referenced to a plane wave propagating in vacuum.
Since the plasma dispersion scales with $\lambda^2$ even a reasonably small ionization leads to a considerable distortion of the THz field.
As the dominant part of the ionization happens when the IR field is present and especially around the peak of the IR pulse, the part of the THz pulse after this peak propagates through a medium with much higher electron density than the leading edge, therefore the effect of plasma dispersion is also considerably higher.
This effect can be observed by comparing cases with weaker and stronger IR fields i.e. lower and higher ionization.
When a 5.2 fs long IR pulse is used with $6*10^{14}$ W/cm$^2$ peak intensity (causing 4.2\% ionization on axis) the main cause of the distortions is the short Rayleigh range, however with a peak intensity of 10$^{15}$ W/cm$^2$ the total ionization is 21.7\% and the trailing edge of the THz pulse suffers from the effects of plasma dispersion.
This can be seen in \autoref{Fig:01} where the initial field at the focus and the propagated fields at the exit of the interaction region (1 mm) are compared showing that the plasma dispersion introduces a significant blue-shift and loss of pulse energy during propagation. 
With longer, 8 fs pulses the ionization rate at $6*10^{14}$ W/cm$^2$ is still just 5.8\% leaving the Gouy phase shift the main cause of distortion.

\subsection{Single atom response and attosecond pulses in the near field}
\label{sec:singleatom}

The effect of THz fields on high-order harmonic generation has been investigated at the single atom level using the classical model \cite{milosevic}, semi-classical model \cite{songsong}, using zero range potential calculations \cite{minqi}, SFA together with saddle point approximation \cite{varjuthz} and by solving the time dependent Schr\"{o}dinger equation for a model atom \cite{milosevic,hong,yang,xiaPhysRevA.81.043420}.
In this section we summarize these results briefly and present the single atom response calculated with our parameters by numerically integrating the Lewenstein integral (\autoref{eq:lew}).
The propagated, and radially integrated intensities of the harmonic bursts at the exit of the interaction region (i.e. in the near field), hence including all the macroscopic effects, are also presented here.

Addition of a static electric or THz field to the generating laser pulse breaks the half-cycle symmetry of the HHG process, and leads to the appearance of both odd and even harmonics in the spectrum \cite{milosevic,hong}.
The additional low frequency field can prevent the closing of some trajectories and if the laser pulse is short enough a broad supercontinuum may appear in the spectra, which leads to the generation of single attosecond pulses.
As reported in \cite{yang} the origin of this supercontinuum is radiation from recombining electrons passing through short trajectories, the THz field suppressing the emission from the corresponding long ones.
It is also observed that with a strong negative chirp of the laser pulse, the width of this supercontinuum may be further increased to obtain a flat spectrum as wide as 700 harmonic orders \cite{yang}.
However laser pulses and static electric fields used by Xiang et al. \cite{yang} are not yet available, hence we focused our study on longer laser pulses without chirp and THz pulses instead of static electric fields.

\begin{figure}[htb!]%
	\subfigure{
	\includegraphics{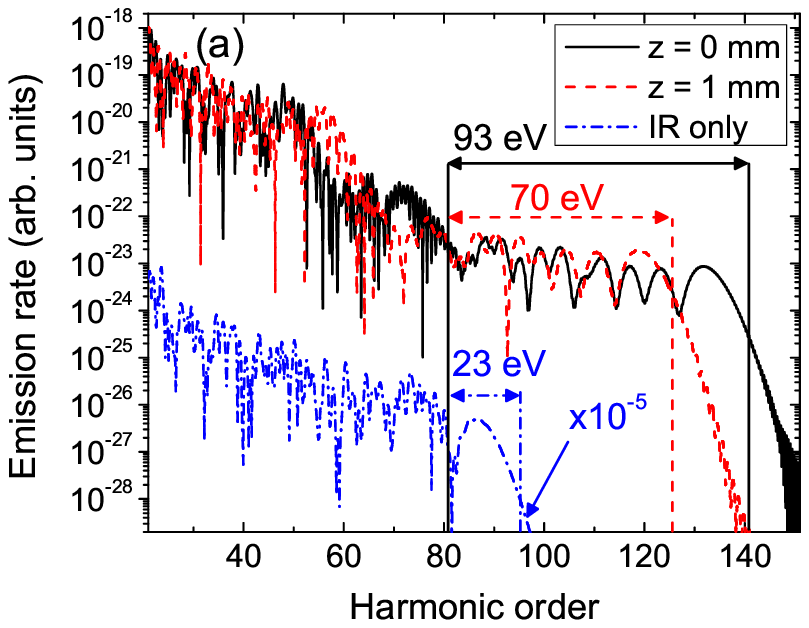}%
	\label{Fig:spec5fs}}
	\subfigure{
	\includegraphics{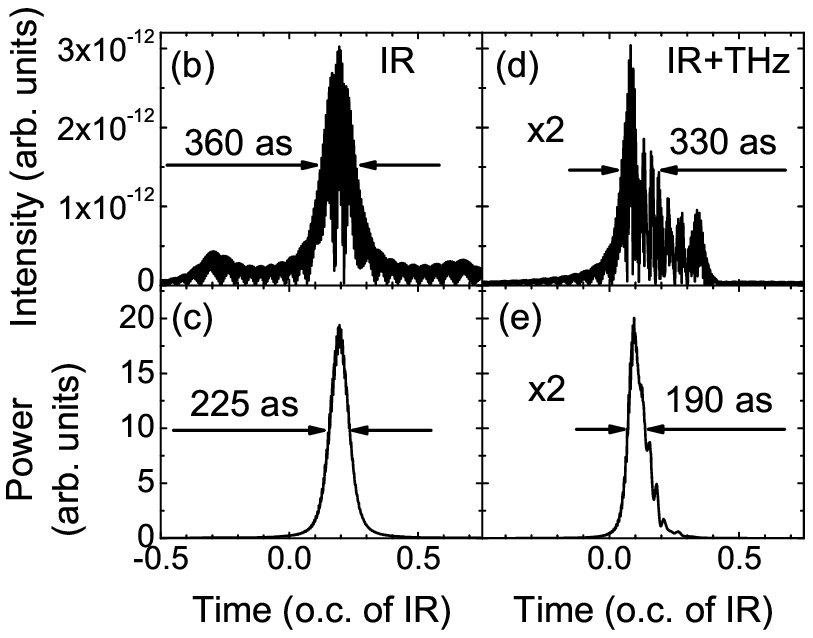}%
	\label{Fig:pulse5fs}}
\caption{(Color online) (a) Harmonic spectra from a single atom, generated at the beginning (black solid line) and at the end (red dashed line) of the cell obtained by using a 5.2 fs IR laser pulse combined with the 100 MV/cm THz pulse. The spectrum on the bottom (blues dash-dotted line) shifted 5 orders of magnitude downward shows the generated spectrum by the same IR pulse but without the THz field. (b) The resulting attosecond pulse from the single atom response for the case without the THz field, and (c) the propagated and radially integrated harmonic field intensity at the exit of the interaction region (also in the case with only the IR field present).
Figure (d) shows the attosecond pulse from the single atom response and (e) the propagated and radially integrated harmonic field intensity (resulting power) at the exit of the interaction region for the case with the THz pulse.
The attosecond bursts were synthesized by selecting harmonic orders $\geq$81 from the harmonic spectra.
}%
\label{Fig:sa5fs}%
\end{figure}

SAPs can be obtained without the need of advanced gating techniques and control fields, 'just' by using adequate spectral filtering, and sufficiently short laser pulses.
For example, a 5.2 fs laser field with $6*10^{14}$ W/cm$^{2}$ peak intensity, generates a harmonic spectrum with a cutoff around the 90$^{th}$ order (see \autoref{Fig:spec5fs}).
By selection of harmonics $\geq$81 an isolated attosecond pulse can be obtained at the single atom level (see \autoref{Fig:pulse5fs}).
For our conditions of cell length, pressure and ionization level, propagation does not distort the laser field. 
Additionally, the cell starts at the focus and we select cutoff harmonics, so the conditions for good phase matching are met \cite{salierePRL1995, balcoupra55}, thus a Gaussian-like pulse is observed in the near field.
The duration of this pulse is 360 as produced at single atom level which becomes 225 as in the near field. 
We note here, that attosecond pulses synthesized from the cutoff posses no chirp \cite{varjujmo,krauszrevmod}.
Increasing the spectral window leads to the appearance of two additional attosecond pulses at half-cycle delay before and after the central one.

On addition of the 100 MV/cm THz field, the spectrum is reshaped to a two-plateau structure, one ending at around order 81 and the other extending its cutoff to harmonic 135; the calculations reveal that the two plateaus are generated in consecutive half-cycles. 
As demonstrated by the trajectory analysis, the second plateau of the spectrum contains only emissions from a pair of a short and a long trajectory, being emitted in a specific optical half-cycle. 
At the end of the interaction region the single atom cutoff is reduced to the 125$^{th}$ harmonic order as a result of the distortions (reducing amplitude, and phase shift) of the THz field during propagation.

Using the same spectral filter as before ($\geq$81$^{st}$ harmonic), a single burst is obtained with full-width at half-maximum of 330 as. 
Although the spectrum is much wider than in the IR alone case, the presence of both trajectories and their phase modulation (chirp) produces a pulse duration comparable to that in the IR only case. 
At the exit of the interaction region macroscopic effects reduce the contribution from the long trajectory components, and hence the duration of the SAP to just 190 as.

\begin{figure}[htb!]%
	\subfigure{
	\includegraphics{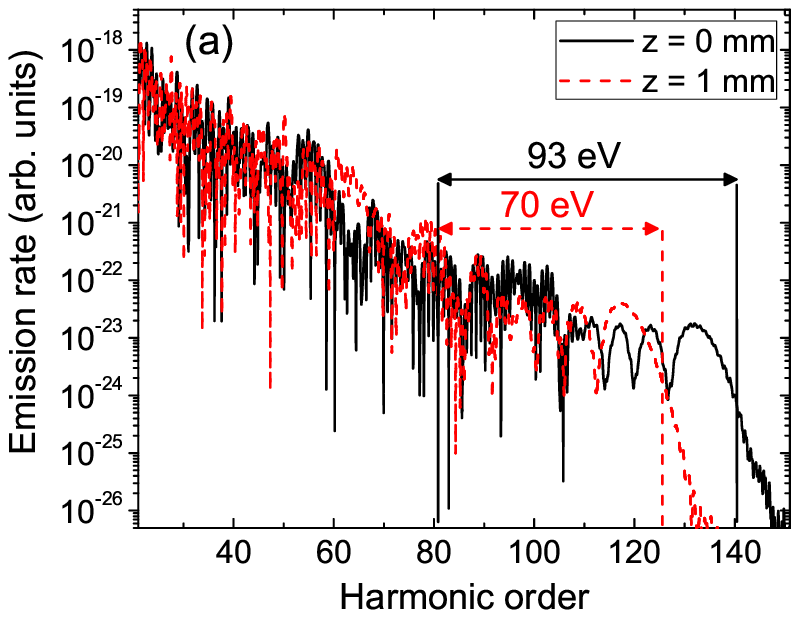}%
	\label{Fig:spec8fs}}
	\subfigure{
	\includegraphics{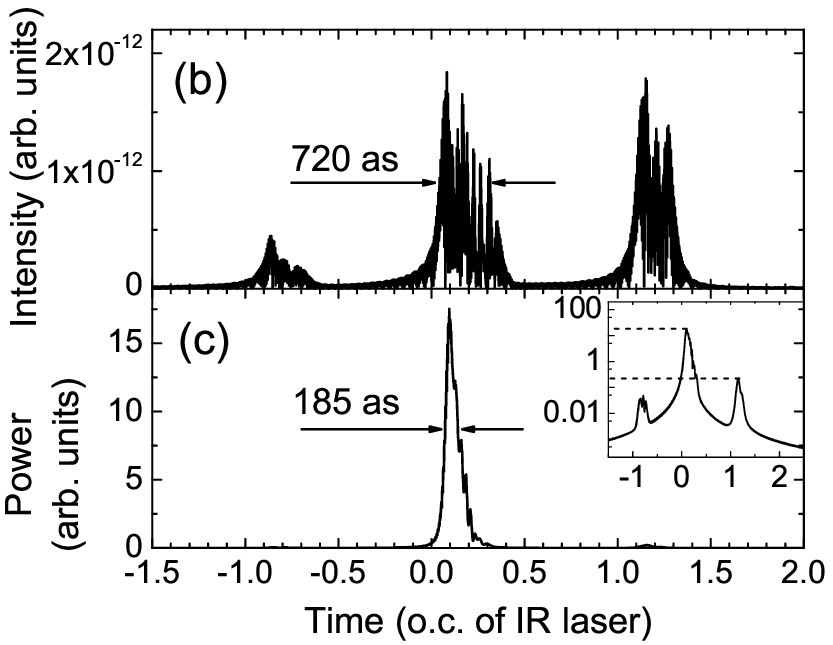}%
	\label{Fig:pulse8fs}}
\caption{(Color online) (a) Harmonic spectra from a single atom, generated at the beginning and at the end of the cell obtained by using an 8 fs IR laser pulse combined with the 100 MV/cm THz pulse. (b) The resulting attosecond pulse from single atom response obtained by selecting harmonics $\geq$81, and (c) the propagated and radially integrated harmonic field intensity at the exit of the interaction region showing a clean SAP. The inset in logarithmic scale shows a contrast of almost 10$^{2}$ between the main pulse and the second most powerful one.}%
\label{Fig:sa8fs}%
\end{figure}

When a longer laser pulse is used (8 fs, with the same $6*10^{14}$ W/cm$^2$ peak intensity, containing 0.45 mJ energy), the part of the spectra from orders 80 to 100 becomes more modulated, suggesting the interference of more trajectories.
Using the same spectral filter as before ($\geq$81), we obtain three distinct attosecond pulses generated at the single-atom level. 
However only the central one survives the propagation, and can be seen at the exit of the interaction region having considerable intensity.
The details of the macroscopic effects responsible for the cleaning and shortening of the attosecond bursts are discussed in the following subsection. 

\subsection{Phase matching effects}
\label{sec:pm}

\begin{figure}[htb!]%
	\subfigure{
	\includegraphics{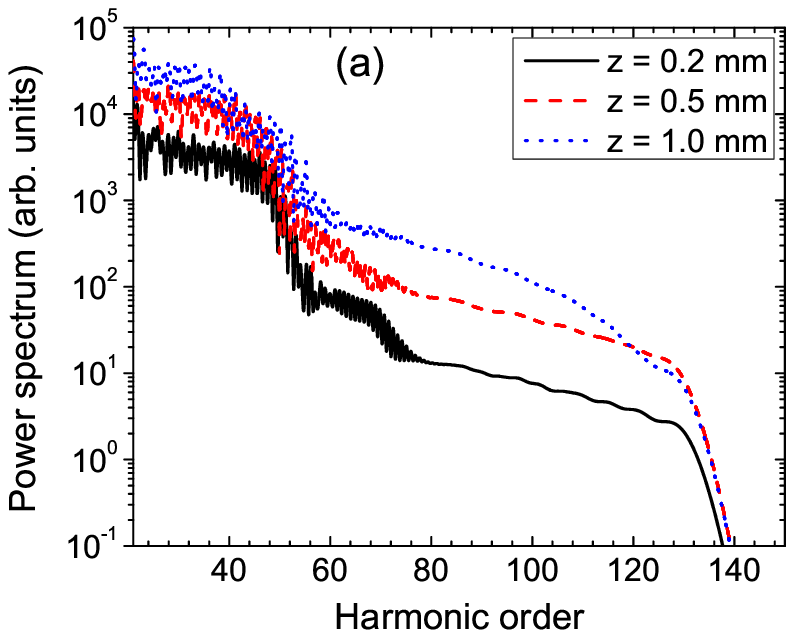}%
	\label{Fig:dnlza}}
	\subfigure{
	\includegraphics{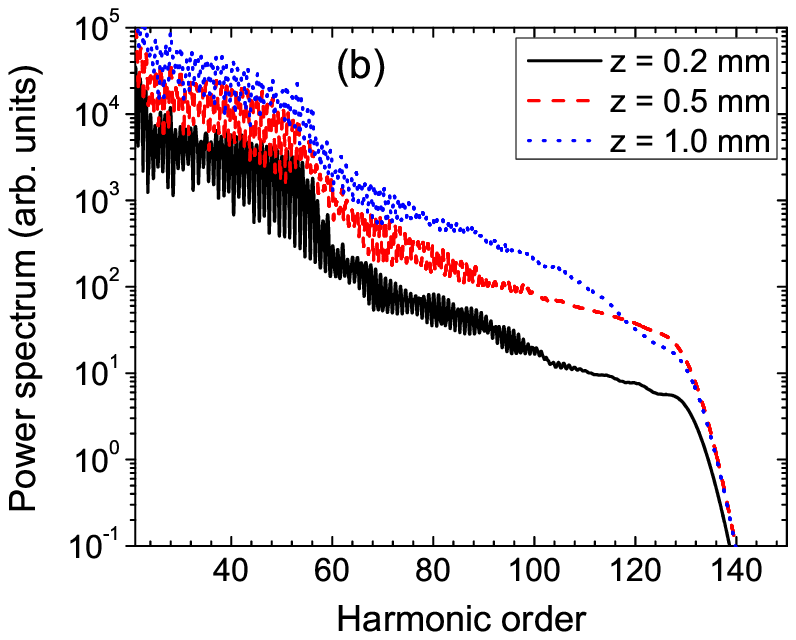}%
	\label{Fig:dnlzb}}
\caption{(Color online) Radially integrated spectral intensity of the propagated field after different lengths of the gas cell for the (a) 5.2 fs, and the (b) 8 fs laser pulses combined with the 100 MV/cm THz pulse. The spectra shown for z=1 mm correspond to the temporal shapes shown previously in Fig. 2(e) and Fig. 3(c), respectively.}%
\label{Fig:nspec}%
\end{figure}

Due to the distortion of the THz field during propagation, the harmonic generation conditions vary substantially along the axial coordinate.
The selection of the central burst seen in \autoref{Fig:pulse8fs} in the 8 fs case also suggests that phase matching promotes only a distinct class of trajectories, and the others are eliminated because of destructive interference.
In order to investigate phase matching of the single atom spectra during propagation, the propagated and radially integrated harmonic intensities (power density spectra) are plotted at different axial ($z$) coordinates for the $6*10^{14}$ W/cm$^2$ case, see \autoref{Fig:nspec}.
For harmonics up to the order of 115 the spectral power increases with propagation distance, suggesting good phase matching for the whole length of the cell. 
On the other hand for the highest harmonics there is no increase for the second part of the cell.

\begin{figure*}[htb!]%
 	\includegraphics{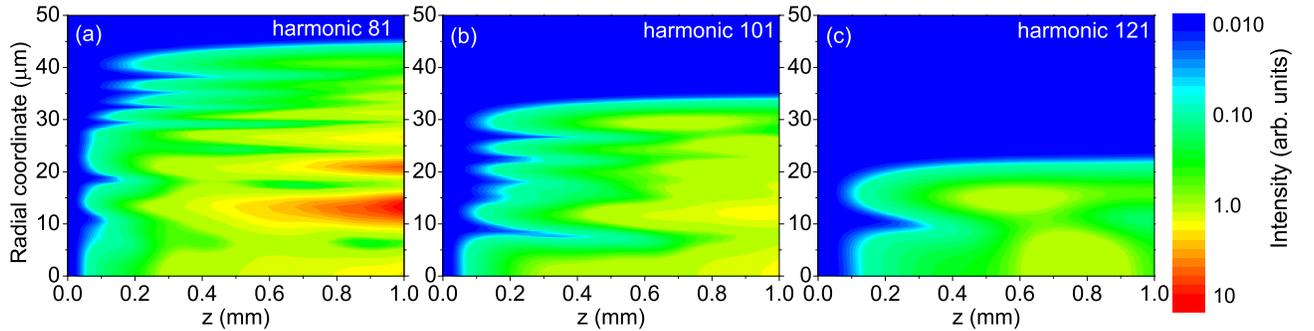}%
\caption{(Color online) Spectral intensity of the propagated harmonic field as a function of radial and axial (z) coordinate for harmonic 81 (a), 101 (b) and 121 (c) calculated for generation with the 8 fs laser pulse.}%
\label{Fig:dnlrz}%
\end{figure*}

Analyzing the spatial structure of different harmonics along the $r$ and $z$ axes, one can see where good phase matching conditions are fulfilled for a specific harmonic (see \autoref{Fig:dnlrz}).
These maps show for example that the intensities of harmonics from order 81 to 101, which belong to the lower part of the plateau, undergo a constant increase along the propagation direction with the best rate slightly off axis. 
Harmonic 121 is phase matched close to the beam axis but only in the first part of the medium, while after \~800 $\mu$m of propagation the field intensity decreases. 
The reason for this decrease is the phase mismatch of the specific harmonic and not the reabsorption of harmonic radiation by the medium. 
This claim is supported by the fact that the cutoff on axis is still slightly above harmonic 121 at the exit of the interaction region and the absorption length is larger than 20 mm for this frequency and in these conditions.

We can conclude that the spectral power density in the lower and middle part of the plateau increases through the propagation, but this analysis is not conclusive on whether phase matching conditions promote certain sets of trajectories. 
This would have a strong effect on the shape, duration and chirp of the resulting pulses.

It is well documented that in HHG with only the IR laser pulse, mainly short trajectory components survive the propagation in a long interaction region, especially after focus, however, in gas cells only a few millimeters long, the contribution of long trajectories to the final pulse might still be significant \cite{gaardejphysb} in specific conditions.
The short- or long-trajectory origin of the resulting burst is important since it defines the temporal, and may alter the spatial properties of the resulting attosecond pulses.
Attosecond bursts from short/long trajectories have positive/negative chirp \cite{varjulaserphys}.
Likewise, radiation generated from short/long trajectories usually has a lower/higher divergence \cite{platonenko_spacialfilter,mairessePRL2004}, making radiation from short trajectories more suitable for applications.
Therefore it is important to investigate whether short or long trajectory components survive the propagation through a 1 mm long gas cell.

We mentioned in the previous subsection that the high frequency part of the single dipole spectrum ($\geq$81 harmonic order) consist of two sets of trajectories.
This is visible in the temporal profile of the attosecond burst - obtained by Fourier transforming the filtered complex spectra - where the presence of short and long trajectories is observable as two separate peaks at a delay less than half IR optical cycle.

\begin{figure*}[htb!]%
	\subfigure{
	\includegraphics{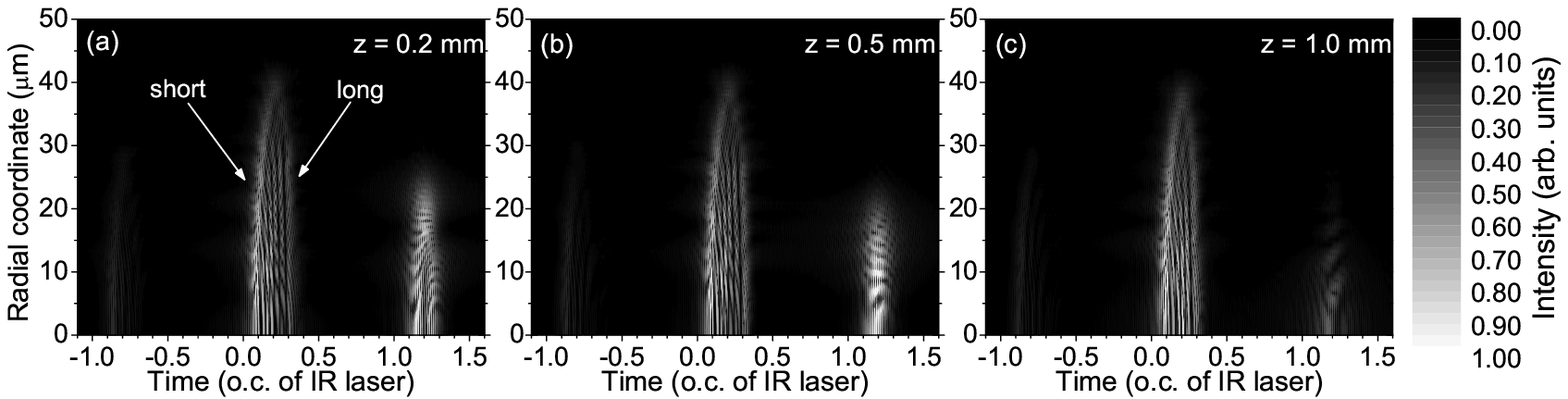} %
	\label{Fig:drta}}
	\subfigure{
	\includegraphics{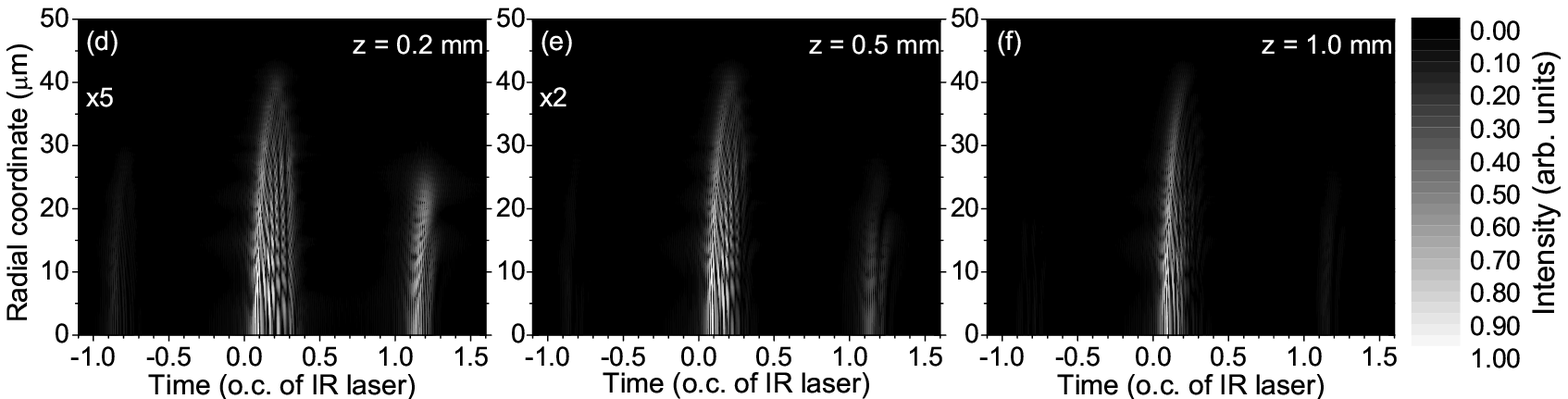} %
	\label{Fig:drtb}}
\caption{(a),(b),(c) Harmonic field intensity of the generated single atom emission at different axial (z) and radial (r) coordinates, and (d),(e),(f) the  intensity of the propagated field at the same coordinates.}%
\label{Fig:drt}%
\end{figure*}

We present in the upper row of \autoref{Fig:drt} the $(t,r)$ map of the single dipole bursts produced by an 8 fs pulse at $6*10^{14}$ W/cm$^2$ at different propagation distances in the gas cell.
We observe that the short- and long- trajectory components merge into the cutoff while the radial coordinate increases(corresponding to decreasing field strength).
We indicate in \autoref{Fig:drta} the short- and long-trajectory classes, as deduced from the trajectory analysis.
As the laser and THz intensity decreases along the propagation direction (due to beam divergence and plasma defocusing), there is no change in the generation of the central attosecond burst, whereas the post pulse strength decreases along the cell.
This can be attributed to the accentuated decrease of the THz field amplitude and phase shift during propagation in the ionized medium, which decreases the cutoff at that specific half-cycle to near the lower limit of the spectral domain from which the harmonic burst is synthesized.
The inner structure appearing inside the attosecond bursts is attributed to interference from different electron trajectories.

The same set of plots has been produced for the propagated harmonic fields (bottom row of \autoref{Fig:drt}) to study the effect of phase matching.
Two bursts separated by an IR cycle are observed, similar to the single atom results above. 
While the main burst is building up during propagation, the strength of the post pulse is decreasing suggesting unfavored phase matching.
Most importantly we would like to point out that, although long trajectory components are generated at any axial coordinate along the cell (top graphs in \autoref{Fig:drt}), they gradually disappear from the propagated field, suggesting phase mismatch for these emissions.
We conclude that harmonic emissions from electrons traveling along short trajectories are well phase matched during propagation, while those from long trajectories are gradually eliminated by destructive interference.

\section{Discussion}
\label{sec:discussion}

In this section we review the effect of several parameters on SAP production, the reference parameters being the ones described in \autoref{sec:setup} in particular we remind the reader that we used an 8 fs laser pulse with a peak intensity of $6*10^{14}$ W/cm$^2$, focused to a 76 $\mu$m beam waist into a 1 mm long gas cell with 15 Torr pressure. The peak amplitude of the THz field was 100 MV/cm.
By selecting harmonic orders $\geq$81 this configuration yielded a SAP with 185 as duration and a contrast ratio of 85:1.
In each of the following subsections the effects of varying one parameter is discussed, except in \autoref{subsec:delay} where the delay between the IR and THz pulses and the length of the interaction region are discussed together.

\subsection{Laser pulse duration}

For the duration of the IR pulse, it seems that the shorter the pulse the better for the positive effects of the THz field, although we did not explore extremal cases like sub-cycle pulses.
Using 5.2 fs laser pulses the continuum part in the near field spectrum starts at the 71$^{st}$ harmonic order and therefore a much wider spectral range can be used for SAP production.
However, because of the chirp of the resulting harmonic pulse, the wider spectral range does not decrease the duration of the obtained SAP, this can be balanced by eliminating cutoff harmonics by spectral filtering.
For example, for a 5.2 fs pulse and selecting harmonic orders 81-115 a 160 as pulse is predicted.
To take advantage of the broad bandwidth an XUV pulse shaping method needs to be implemented \cite{broadbandPulseShaping,xuvchirpmirror} to obtain transform limited pulses.
With a suitable chirp compensation technique the IR only SAP duration of 225 as is reduced to 210 as, while the 185 as pulse generated in the presence of the THz field is reduced to just $\approx$50 as.

\begin{figure*}[htb!]%
	\includegraphics{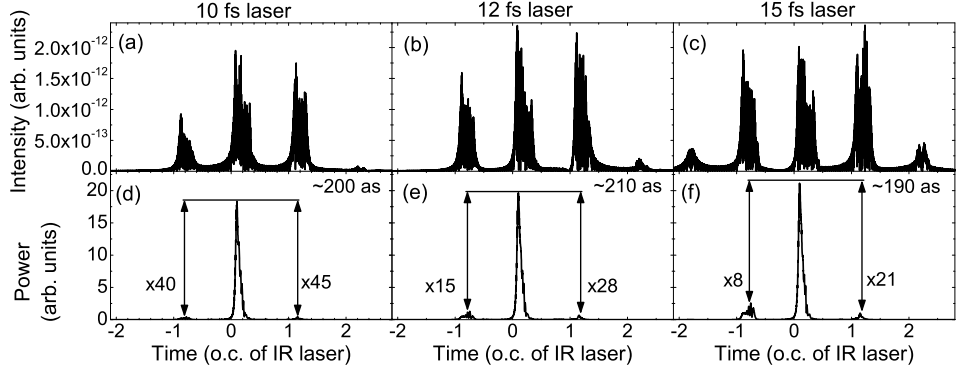} %
\caption{Generated harmonic bursts at the single atom level (top row) and the propagated and radially integrated harmonic field intensities at the exit of the interaction region (bottom row) for different generating laser pulses having 10 fs (a) and (d), 12 fs (b) and (e), and 15 fs duration (c) and (f), assisted by the THz field.}%
\label{Fig:pulsed}%
\end{figure*}

Longer laser pulses may be used for SAP production, but this affects the contrast.
Using the same spectral filtering (harmonics $\geq$81) and 10 fs laser pulse the contrast is decreased to 40:1 and this ratio is further decreased to 15:1 with 12 fs, and to 8:1 with 15 fs pulses. 
By increasing the lower limit of the spectral filter the contrast can be slightly increased at the cost of reduced power of the main pulse.

\subsection{Gas pressure}

\begin{figure*}[htb!]%
	\includegraphics{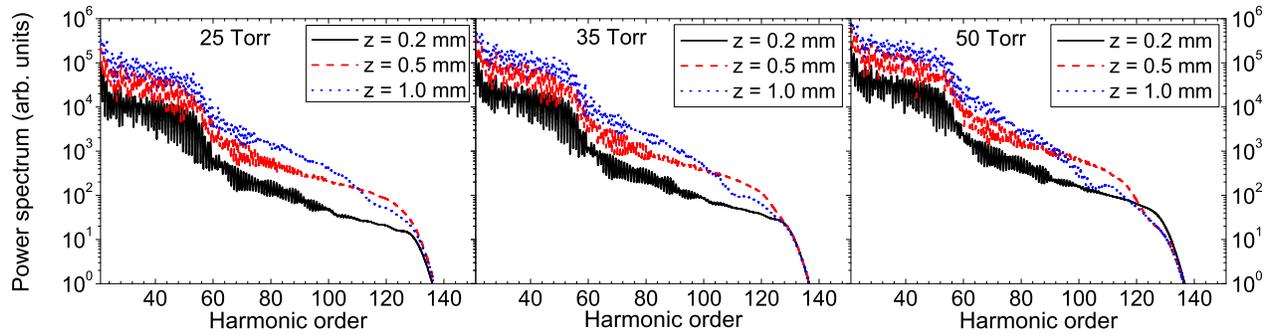} %
\caption{(Color online) Radially integrated spectral intensity of the propagated field after different lengths of the gas cell for 25, 35 and 50 Torr gas pressure showing the detrimental effect of high gas pressure on the phase matching of cutoff harmonics.}%
\label{Fig:specp}%
\end{figure*}

The increase of the gas pressure has a detrimental effect for the phase matching in the high frequency range thus narrowing the spectral range available for SAP generation. 
However the harmonics involved in SAP formation is closer to the cutoff so the resulting SAP duration is decreased.
We found that the optimum pressure in terms of contrast ratio is around 25 Torr.
At this pressure the contrast is increased to 170:1, the peak power of the SAP is doubled and its duration is reduced to 165 as when the same spectral filtering as before is used.
On further increase of the gas pressure the pulse duration is further decreased to 140 and 130 as with 35 and 50 Torr gas pressure respectively; however, this also results in a decrease in contrast to 130:1 and 35:1.

\subsection{Delay between THz and IR pulses and optimal cell length}
\label{subsec:delay}

Another important parameter is the length of the interaction region, the optimal value of which is limited by phase matching conditions and reabsorption.
With the base configuration where the Rayleigh range of the THz pulse is 2.29 mm the most powerful SAP can be obtained with a 1.3 mm long cell.
In this case the near field SAP duration is reduced to 150 as and the contrast is increased to 130:1.

By change in the delay from -2.0 fs (IR pulse behind) to 1.5 fs (IR pulse ahead) the near field cutoff is increased from harmonic 85 to 125. 
When the IR is 1.5 fs ahead, the near field spectral power density for harmonics in the cutoff region is increased 4.5 times compared to the case without delay (using a 1 mm cell in both cases).
However, by putting the IR pulse ahead of the THz by 1.5 fs, the optimal cell length (in terms of harmonic pulse power) is also increased from 1.3 to 2 mm. 
In this case the harmonic pulse's peak power is doubled, the duration of the obtained SAP is 150 as, and the contrast is 75:1.

\subsection{Laser pulse energy}

By increasing IR pulse energy we could still generate SAP and favorable phase matching for the short trajectories.
Of course, due to the increased cutoff we also adjusted the spectral filtering for an optimum SAP generation.
With $8*10^{14}$ W/cm$^2$ peak intensity for the 8 fs pulse, the result is 16.4\% peak ionization and a 165 as SAP having a contrast of 230:1 when harmonics $\geq$101 are selected.
By further increasing peak intensity to 10$^{15}$ W/cm$^2$ the resulting ionization is 32\%, (compared to 21.7\% in \autoref{Fig:01b} with a 5.2 fs laser pulse having the same peak intensity) which affects the THz field propagation in its trailing edge.
The gating effect of the THz pulse is still present, and a SAP with 150 as duration and a contrast of 100:1 is obtained in the near field by selecting harmonics $\geq$111. 
In this case the single atom cutoff is at harmonic order 191, although by the end of the 1 mm gas cell it decreases to order 142, it is still higher than in the case without the THz field with cutoffs at harmonic orders 136 and 121, respectively.
Without the THz field, by use of the same spectral filtering two almost identical pulses are obtained (concerning their duration and peak power) with half the IR optical cycle delay between them, and a third one having much lower peak power is also observable.

\section{Conclusions}
\label{sec:conclusions}

Macroscopic effects for high-order harmonic generation in the presence of strong THz fields were studied.
Our results show that the cutoff order for harmonic radiation is increased from 90 to 135.
Furthermore the available bandwidth for SAP generation is increased from 23 eV to 93 eV.
We also show that even in cases when longer laser pulses are used (8, 10 or 12 fs) and the single atom response yields multiple attosecond pulses, propagation effects can eliminate the contribution from certain sets of trajectories, yielding a single attosecond pulse at the exit of the cell (with contrast ratio of 170:1, 40:1, and 15:1, respectively).
The long-trajectory components are also cleaned from the surviving pulse during propagation resulting in an effective decrease of pulse duration, making the technique promising for obtaining a reliable source of short, isolated attosecond pulses with good contrast and low divergence. 
By carefully adjustment of the parameters, such as gas pressure and peak intensity of the laser pulse, 130 as pulses are produced using adequate spectral filtering in a straightforward manner, instead of the 225 as pulse obtained using only the IR pulse.
Due to the large bandwidth the transform limited duration of these pulses is reduced from 210 to just $\approx$50 as in the presence of the THz field.

\acknowledgments
The project was supported by the European Community's Seventh Framework Programme under contract ITN-2008-238362 (ATTOFEL).
KV and PD acknowledge the support of the Bolyai Postdoctoral Fellowship. KV is also grateful for the support of NKTH-OTKA (\#74250).
JAF and JH acknowledge support from Hungarian Scientific Research Fund (OTKA), grant numbers 76101 and 78262, and from `Science, Please! Research Team on Innovation' (SROP-4.2.2/08/1/2008-0011). 
GyF acknowledges support from Hungarian Scientific Research Fund (OTKA), grant number 73728.
KK is grateful for the support of TAMOP-4.2.1/B-09/1/KONV-2010-0005 financed by the European Union and the European Regional Fund.


%

\end{document}